\documentclass{pasj01}

\usepackage{natbib}
\usepackage{time}
\usepackage{color}

\usepackage{mathpazo}
\usepackage[T1]{fontenc}

\Received{yyyy/mm/dd}
\Accepted{yyyy/mm/dd}
\Published{yyyy/mm/dd}

\newcommand{\eg}{e.g., }
\newcommand{\ie}{i.e., }

\newcommand{\Msun}{M_{\odot}}

\newcommand{\Ye}{Y_{\rm e}}
\def\gsim{\mathrel{\rlap{\lower 4pt \hbox{\hskip 1pt $\sim$}}\raise 1pt
\hbox {$>$}}}
\def\lsim{\mathrel{\rlap{\lower 4pt \hbox{\hskip 1pt $\sim$}}\raise 1pt
\hbox {$<$}}}

\begin{document} 

\title{Kilonova from post-merger ejecta as an optical and near-infrared counterpart of GW170817}

\author{Masaomi \textsc{Tanaka}\altaffilmark{1}}
\author{Yousuke \textsc{Utsumi}\altaffilmark{2}}
\author{Paolo \textsc{A. Mazzali}\altaffilmark{3,4}}
\author{Nozomu \textsc{Tominaga}\altaffilmark{5,6}}
\author{Michitoshi \textsc{Yoshida}\altaffilmark{7}}
\author{Yuichiro \textsc{Sekiguchi}\altaffilmark{8}}
\author{Tomoki \textsc{Morokuma}\altaffilmark{9}}
\author{Kentaro \textsc{Motohara}\altaffilmark{9}}
\author{Kouji \textsc{Ohta}\altaffilmark{10}}
\author{Koji S. \textsc{Kawabata}\altaffilmark{2}}

\author{Fumio \textsc{Abe}\altaffilmark{11}}
\author{Kentaro \textsc{Aoki}\altaffilmark{7}}
\author{Yuichiro \textsc{Asakura}\altaffilmark{11,$\dagger$}}
\author{Stefan \textsc{Baar}\altaffilmark{12}}
\author{Sudhanshu \textsc{Barway}\altaffilmark{13}}
\author{Ian A. \textsc{Bond}\altaffilmark{14}}
\author{Mamoru \textsc{Doi}\altaffilmark{9}}
\author{Takuya \textsc{Fujiyoshi}\altaffilmark{7}}
\author{Hisanori \textsc{Furusawa}\altaffilmark{1}}
\author{Satoshi \textsc{Honda}\altaffilmark{12}}

\author{Yoichi \textsc{Itoh}\altaffilmark{12}}
\author{Miho \textsc{Kawabata}\altaffilmark{15}}
\author{Nobuyuki \textsc{Kawai}\altaffilmark{16}}
\author{Ji Hoon \textsc{Kim}\altaffilmark{7}}
\author{Chien-Hsiu \textsc{Lee}\altaffilmark{7}}
\author{Shota \textsc{Miyazaki}\altaffilmark{17}}
\author{Kumiko \textsc{Morihana}\altaffilmark{12,18}}
\author{Hiroki \textsc{Nagashima}\altaffilmark{15}}
\author{Takahiro \textsc{Nagayama}\altaffilmark{19}}
\author{Tatsuya \textsc{Nakaoka}\altaffilmark{15}}

\author{Fumiaki \textsc{Nakata}\altaffilmark{7}}
\author{Ryou \textsc{Ohsawa}\altaffilmark{9}}
\author{Tomohito \textsc{Ohshima}\altaffilmark{12}}
\author{Hirofumi \textsc{Okita}\altaffilmark{7}}
\author{Tomoki \textsc{Saito}\altaffilmark{12}}
\author{Takahiro \textsc{Sumi}\altaffilmark{17}}
\author{Akito \textsc{Tajitsu}\altaffilmark{7}}
\author{Jun \textsc{Takahashi}\altaffilmark{12}}
\author{Masaki \textsc{Takayama}\altaffilmark{12}}
\author{Yoichi \textsc{Tamura}\altaffilmark{20}}

\author{Ichi \textsc{Tanaka}\altaffilmark{7}}
\author{Tsuyoshi \textsc{Terai}\altaffilmark{7}}
\author{Paul J. \textsc{Tristram}\altaffilmark{21}}
\author{Naoki \textsc{Yasuda}\altaffilmark{6}}
\author{Tetsuya \textsc{Zenko}\altaffilmark{10}}

\altaffiltext{1}{National Astronomical Observatory of Japan, 2-21-1 Osawa, Mitaka, Tokyo 181-8588, Japan}
\altaffiltext{2}{Hiroshima Astrophysical Science Center, Hiroshima University, 1-3-1 Kagamiyama, Higashi-Hiroshima, Hiroshima, 739-8526, Japan}
\altaffiltext{3}{Astrophysics Research Institute, Liverpool John Moores University, IC2, 134 Brownlow Hill, Liverpool L3 5RF, UK}
\altaffiltext{4}{Max-Planck-Institut f\"ur Astrophysik, Karl-Schwarzschild-Str. 1, D-85748 Garching bei M\"unchen, Germany}
\altaffiltext{5}{Department of Physics, Faculty of Science and Engineering, Konan University, 8-9-1 Okamoto, Kobe, Hyogo 658-8501, Japan}
\altaffiltext{6}{Kavli Institute for the Physics and Mathematics of the Universe (WPI), The University of Tokyo Institutes for Advanced Study, The University of Tokyo, 5-1-5 Kashiwa, Chiba 277-8583, Japan}
\altaffiltext{7}{Subaru Telescope, National Astronomical Observatory of Japan, 650 North A'ohoku Place, Hilo, HI 96720, USA}
\altaffiltext{8}{Department of Physics, Toho University, Funabashi, Chiba 274-8510, Japan}
\altaffiltext{9}{Institute of Astronomy, Graduate School of Science, The University of Tokyo, 2-21-1 Osawa, Mitaka, Tokyo 181-0015, Japan}
\altaffiltext{10}{Department of Astronomy, Kyoto University, Kitashirakawa-Oiwake-cho, Sakyo-ku, Kyoto,  606-8502, Japan}

\altaffiltext{11}{Institute for Space-Earth Environmental Research, Nagoya University, Furo-cho, Chikusa, Nagoya, Aichi 464-8601, Japan}
\altaffiltext{12}{Nishi-Harima Astronomical Observatory, Center for Astronomy, University of Hyogo, 407-2, Nishigaichi, Sayo, Hyogo 679-5313, Japan}
\altaffiltext{13}{South African Astronomical Observatory, PO Box 9, 7935 Observatory, Cape Town, South Africa}
\altaffiltext{14}{Institute for Natural and Mathematical Sciences, Massey University, Private Bag 102904 North Shore Mail Centre, Auckland 0745, New Zealand}
\altaffiltext{15}{Department of Physical Science, Hiroshima University, Kagamiyama, Higashi-Hiroshima 739-8526, Japan}
\altaffiltext{16}{Department of Physics, Tokyo Institute of Technology, 2-12-1 Ookayama, Meguro-ku, Tokyo 152-8551}
\altaffiltext{17}{Department of Earth and Space Science, Graduate School of Science, Osaka University, 1-1 Machikaneyama, Toyonake, Osaka 560-0043, Japan}
\altaffiltext{18}{Division of Particle and Astrophysical Science, Graduate School of Science, Nagoya University, Furo-cho, Chikusa-ku, Nagoya, 464-8602, Japan}
\altaffiltext{19}{Graduate School of Science and Engineering, Kagoshima University, 1-21-35, Korimoto, Kagoshima, 890-0065, Japan}
\altaffiltext{20}{Division of Particle and Astrophysical Science, Graduate School of Science, Nagoya University, Furo-cho, Chikusa-ku, Nagoya, 464-8602 Japan}
\altaffiltext{21}{University of Canterbury, Mt John Observatory, PO Box 56, Lake Tekapo 7945, New Zealand}

\altaffiltext{$\dagger$}{Deceased 18 August 2017}


\maketitle

\begin{abstract}
Recent detection of gravitational waves from a neutron star (NS) merger event
GW170817 and identification of an electromagnetic counterpart provide
a unique opportunity to study the physical processes in NS mergers.
To derive properties of ejected material from the NS merger,
we perform radiative transfer simulations of kilonova,
optical and near-infrared emissions powered by radioactive decays
of $r$-process nuclei synthesized in the merger.
We find that the observed near-infrared emission lasting for $> 10$ days
is explained by 0.03 $\Msun$ of ejecta containing lanthanide elements.
However, the blue optical component observed at the initial phases
requires an ejecta component with a relatively high electron fraction ($\Ye$).
We show that both optical and near-infrared emissions are simultaneously
reproduced by the ejecta with a medium $\Ye$ of $\sim 0.25$.
We suggest that a dominant component powering the emission
is post-merger ejecta,
which exhibits that mass ejection after the first dynamical ejection
is quite efficient.
Our results indicate that NS mergers synthesize a wide range of $r$-process
elements and strengthen the hypothesis that NS mergers are
the origin of $r$-process elements in the Universe.
\end{abstract}


\section{Introduction}
The first gravitational wave (GW) detection from a neutron star merger
was successfully made for GW170817 \citep{lvc17}.
The detection of GW170817 triggered electromagnetic (EM) observations
over the entire wavelength range.
Gamma-ray signals were detected about two seconds
after the coalescence \citep{connaughton17,savchenko17}.
Then, a promising optical and near-infrared counterpart was discovered
(\citealt{coulter17,allam17,yang17}, Coulter et al. in prep., Soares-Santos et al. in prep., Valenti et al. in prep.)
within a localization area estimated from the three GW detectors
(Advanced LIGO Livingston/Hanford and Advanced Virgo).
The object, named as SSS17a (or DLT17ck), is located in NGC 4993,
a galaxy at a distance of 40 Mpc.
The object was also detected subsequently
in ultraviolet \citep{evans17}, X-ray \citep{fong17,troja17},
and radio (\citealt{mooley17,corsi17}; Hallinan et al. submitted) wavelengths.

Among various EM signals from NS mergers,
optical and near-infrared emission is in particular of interest.
When two NSs merge, a small part of NS material
($\sim 10^{-3} - 10^{-2} \Msun$) is ejected into interstellar space
\citep[\eg][]{rosswog99,ruffert01,goriely11,hotokezaka13,bauswein13}.
In the ejected material,
rapid neutron capture nucleosynthesis ($r$-process) takes place
\citep[\eg][]{lattimer74,eichler89,korobkin12,wanajo14}.
Then, radioactive decays of $r$-process nuclei powers
EM emission especially in the optical and near-infrared wavelengths
\citep{li98,kulkarni05,metzger10,kasen13,barnes13,tanaka13}.
This emission is called ``kilonova'' or ``macronova''
\citep[see][for reviews]{fernandez16,tanaka16,metzger17}.

NS mergers eject material by several mechanisms.
The first and robust mechanism is a dynamical mass ejection,
which is driven by tidal disruption as well as shock heating.
Properties of the dynamical ejecta depend on the mass ratio of
two NSs and the equation of state for a NS \citep[\eg][]{hotokezaka13}.
After the dynamical ejection, there may also be additional mass ejection
(``post-merger'' ejecta or ``wind'' ejecta) by viscous heating
and neutrino heating \citep[\eg][]{dessart09,fernandez13,perego14,shibata17}.
However, detailed properties of the post-merger ejecta, such as the ejecta
mass, electron fraction ($\Ye$, number of protons per nucleon which controls
the final element abundances), and its spatial distribution, are still
unclear due to difficulties in resolving the turbulence and sophisticated
treatment of neutrinos in numerical simulations
\citep[see \eg][]{price06,kiuchi14,giacomazzo15,siegel17}.

The optical and near-infrared counterpart to GW170817, SSS17a, provides
an excellent opportunity to study the mass ejection mechanisms of
NS mergers and nucleosynthesis in the ejecta.
The observed properties of SSS17a are summarized as follows \citep{utsumi17}.
(1) The optical and near-infrared brightnesses reach
absolute magnitudes of $-15$ or $-16$ mag (AB magnitude).
(2) The optical light curves decline rapidly
while the near-infrared light curves evolve more slowly:
the flux is dominated by near-infrared light at $t > 5$ days after GW170817
(hereafter $t$ denotes the time after GW170817).
(3) The emission at the initial phase ($t < 2$ days) is relatively blue,
dominated by the optical light.
(4) A non-thermal afterglow component is likely to be negligible 
since the optical and near-infrared spectra of SSS17a are dominated
by a thermal component from the initial phases,
at least $t = 1.5$ days \citep{pian17}.
\citet{utsumi17} shows that rapid evolution of SSS17a is
consistent with the expection of kilonova models.
In this paper, we perform radiative transfer simulations of kilonovae,
and study the mass ejection from the NS merger event GW170817.


\begin{figure}
 \begin{center}
   \includegraphics[scale=0.9]{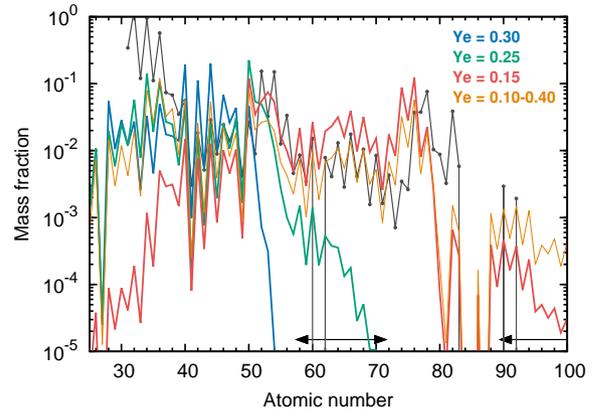}
\end{center}
 \caption{Element abundances at 1 day after the merger.
   The blue, green and red lines show the abundance patterns calculated with
   $\Ye$ = 0.30, 0.25, and 0.15, respectively \citep{wanajo14}.
   The abundance patterns of $\Ye$ = 0.30 and 0.25 approximate
   the post-merger ejecta. The orange line shows the abundance
   by assuming a flat distribution of $\Ye$ from 0.10 to 0.40,
   which depicts the properties of the dynamical ejecta.
   The black line shows the solar abundances.
   The arrows show the lanthanide and actinide elements with high opacities.}
 \label{fig:abundance}
\end{figure}

\begin{figure*}
  \begin{center}
    \begin{tabular}{cc}
      \includegraphics[scale=0.9]{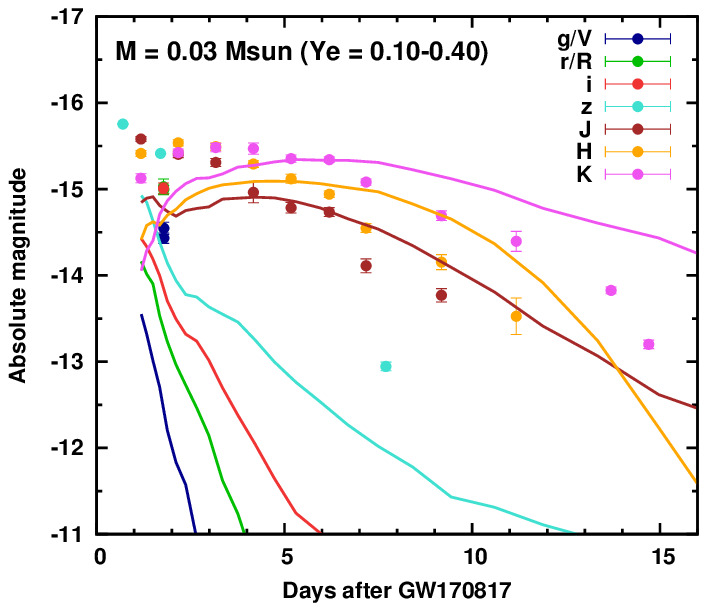}&
      \includegraphics[scale=0.9]{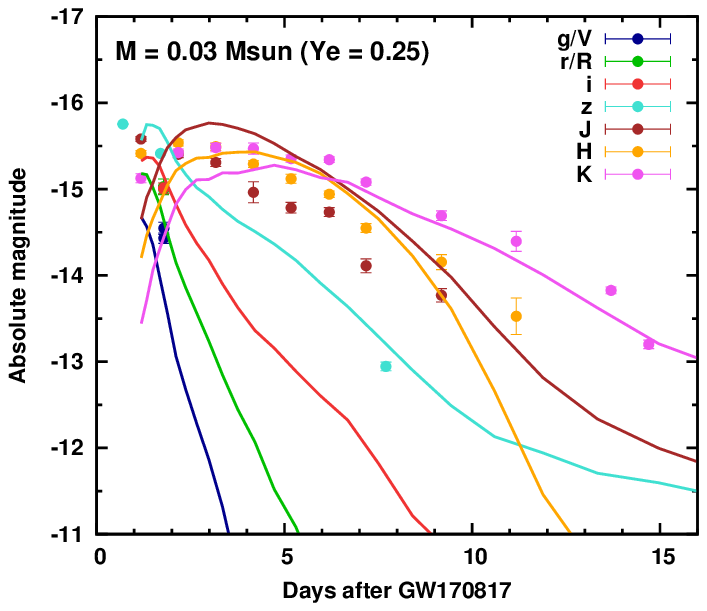}
      \end{tabular}
\end{center}
 \caption{
   Optical and near-infrared light curves of SSS17a compared with
   kilonova models with (left) $\Ye = 0.10-0.40$
   and (right) $\Ye = 0.25$.
   The optical and near-infrared data are taken from \citet{utsumi17}.
   For the observed data, the line of sight extinction of $E(B-V)$ = 0.1 mag
   has been corrected. All the magnitudes are given in AB magnitudes.
 }
 \label{fig:LC}
\end{figure*}

\section{Radiative Transfer Simulations}
\label{sec:simulations}

We perform radiative transfer simulations to calculate light curves
and spectra of kilonovae.
We use a wavelength-dependent radiative transfer code
\citep{tanaka13,tanaka14,tanaka16,tanaka17}.
For a given ejecta model, \ie
a density structure and element abundances,
the code calculates photon transfer by the Monte Carlo method.
For a given $\Ye$ (or distribution of $\Ye$),
the code adopts nuclear heating rates by $r$-process
nucleosynthesis calculations by \citet{wanajo14}.
Then, time-dependent thermalization efficiency is taken into account
by following the analytic methods presented by \citet{barnes16}.

A transfer of optical and near-infrared photons is calculated by
taking into account bound-bound, bound-free, and free-free transitions
and electron scattering.
Among them, the bound-bound transitions have a dominant contribution
in the optical and near-infrared wavelengths.
For the bound-bound transitions, formalism of the expansion opacity
\citep{eastman93,kasen06} is used as in the previous kilonova simulations
\citep{kasen13,barnes13,tanaka13}.
For atomic data, we use a line list by \citet{tanaka17},
which is constructed by the atomic structure calculations for Se ($Z = 34$),
Ru ($Z=44$), Te ($Z=52$), Nd ($Z=60$), and Er ($Z=68$)
and supplemented by Kurucz's line list for $Z < 32$ \citep{kurucz95}.
Note that since the atomic data include only up to doubly ionized ions,
our calculations are applicable only after $t > 0.5$ days,
when the temperature is low enough.
The number density of each ion is calculated under an assumption
of local thermodynamic equilibrium, and populations of excited levels
are calculated by assuming the Boltzmann distribution.

For a density distribution in the ejecta,
we adopt a simple power-law form ($r^{-3}$) from $v = 0.05c$ to $0.2c$,
which gives the average velocity of $\langle v \rangle = 0.1c$,
as a representative case \citep{metzger10,metzger17}.
We test three different element abundances,
which approximate the dynamical ejecta and post-merger ejecta.
The first case depicts the abundances in the dynamical ejecta.
Numerical relativity simulations of NS mergers predict wide ranges of $\Ye$
in the dynamical ejecta \citep{sekiguchi15,sekiguchi16,radice16,foucart16},
which results in a wide elemental distribution from $Z \sim 30$ to 100.
Such element abundances are shown in the orange line
in Figure \ref{fig:abundance}, which are calculated
by assuming a flat $\Ye$ distribution from 0.10 to 0.40 \citep{wanajo14}.
The second and third cases are for the post-merger ejecta.
Since the element abundances are subject to uncertainties,
we approximately take two representative values of $\Ye$:
high $\Ye$ ($\Ye$ = 0.30, blue line) and
medium $\Ye$ ($\Ye$ = 0.25, green line).
The high $\Ye$ model is completely lanthanide-free
while the medium $\Ye$ model contains a small fraction of lanthanide elements.
For all the models in this paper,
the element distribution in the ejecta is assumed to be spatially homogeneous.
Validity of this assumption is discussed in Section \ref{sec:discussions}.

\begin{figure}
 \begin{center}
  \includegraphics[scale=1.0]{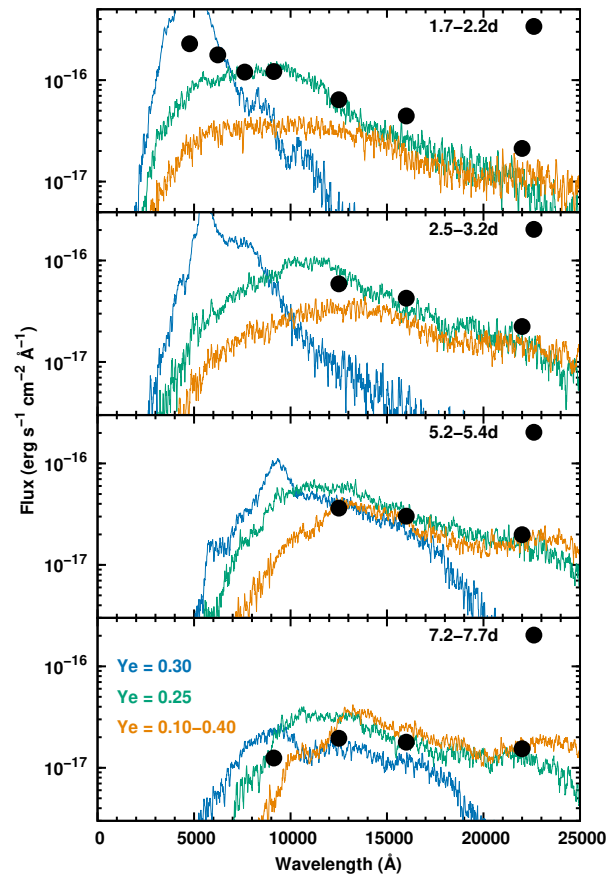}
\end{center}
 \caption{Time evolution of optical and near-infrared spectral energy
   distribution of SSS17a compared with three models.
   The observational data are taken from \citet{utsumi17}.
   All of the three models assume the same ejecta mass ($0.03 \Msun$)
   and the same average velocity ($\langle v \rangle = 0.1c$).
   Orange curves show the model of the dynamical ejecta ($\Ye$ = 0.10-0.40)
   while blue and green curves show the models with the elemental
   abundances calculated with high $\Ye$ ($\Ye$ = 0.30)
   and medium $\Ye$ ($\Ye$ = 0.25), respectively.}
 \label{fig:spec}
\end{figure}

\section{Results}
\label{sec:results}

The left panel of Figure \ref{fig:LC} compares the observed light curves
of SSS17a \citep{utsumi17} and the model with $\Ye = 0.10-0.40$
(the dynamical ejecta model).
We find that the ejecta mass of 0.03 $\Msun$ reasonably
reproduces the near-infrared brightness near the peak.
However, the calculated optical light curves are systematically
fainter than the observations by 1.0-1.5 mag
at the initial phases ($t < 2$ days).
This is due to high optical opacities of lanthanide elements
($Z = 57-71$, \citealt{kasen13,barnes13,tanaka13,fontes17,wollaeger17}).
Because this model has a considerable fraction of lanthanide
elements, the resulting kilonova at the initial phases is
too red compared with the observations.
The faint optical flux is also shown in Figure \ref{fig:spec},
where the spectral energy distribution of SSS17a is compared with
simulated spectra (orange line for $\Ye = 0.10-0.40$).
To explain the optical brightness,
ejecta mass of $\sim 0.06 \Msun$ is required,
although such a model gives too bright near-infrared light curves.

The observed blue emission at the initial phases indicates
a presence of the ejecta with relatively low opacities.
The green curves in Figure \ref{fig:spec} show simulated spectra
of the model with the ejecta mass of $0.03 \Msun$ and $\Ye$ = 0.25.
The overall agreement between the observed spectral energy distribution
and that of the medium $\Ye$ model is satisfactorily well
in both the optical and the near-infrared wavelengths from $t=2$ days to 7 days.
As expected from the good agreement with the spectral energy distribution,
the model with medium $\Ye$ also reproduces the overall properties
of the multi-color light curves (right panel of Figure \ref{fig:LC}).
If the ejecta are completely free from lanthanide elements
($\Ye$ = 0.30, blue lines in Figure \ref{fig:spec}),
the spectra are too blue and do not produce enough flux
in the near-infrared wavelengths ($> 10,000$ \AA) at all the epochs.

\begin{figure}
 \begin{center}
  \includegraphics[scale=1.15]{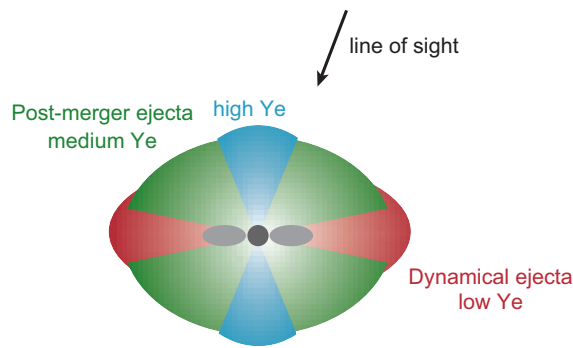}
\end{center}
 \caption{Schematic picture of the ejecta of the NS merger event GW170817.}
 \label{fig:schematic}
\end{figure}

\section{Discussions}
\label{sec:discussions}

A comparison between our radiative transfer simulations
and the observations of SSS17a provides insight
on the ejected material in the NS merger event GW170817.
We show that the observed near-infrared emission
is nicely explained by $0.03 \Msun$ of ejecta
containing lanthanide elements 
($\Ye = 0.10-0.40$ or $\Ye = 0.25$).
However, the model with $\Ye = 0.10-0.40$
does not reproduce the blue optical emission at the initial phases.
On the other hand, if the ejecta are completely lanthanide free ($\Ye = 0.30$),
the emission is too blue compared with the observations.
We find that, as far as a single component model is considered,
the model with $\Ye = 0.25$ containing a small fraction of lanthanide elements
reproduces both optical and near-infrared emissions reasonably well.

What is the origin of such ejecta?
The simulations of the dynamical mass ejection
show that a stronger mass ejection occurs when radii of the NSs are smaller
(\ie when the equation of state of the NSs is soft),
and thus, shock heating is more efficient.
However, a possible maximum mass of the dynamical ejecta is about
0.01 $\Msun$ with currently available equation of states
\citep[\eg][]{hotokezaka13,sekiguchi15,sekiguchi16,radice16}.
An even higher mass ejection might be possible for a merger
with an extreme mass ratio of two NSs.
However, in such cases, a tidally disrupted component
with low $\Ye$ dominates (see the red line in Figure \ref{fig:abundance}
for the abundances with $\Ye$=0.15)
and the emission would become even redder at the initial phases.
By virtue of these facts, it is unlikely that the dynamical ejecta alone
can power entire optical and near-infrared emissions of SSS17a.

We suggest that a kilonova from post-merger ejecta
plays a dominant contribution for SSS17a.
The observed properties are nicely explained
if the entire ejecta is moderately lanthanide-rich
as in the case of $\Ye$ = 0.25.
However, it does not necessarily mean that the ejecta
should have only a single component.
In reality, the ejecta would have an angular distribution of $\Ye$,
having higher $\Ye$ near a polar region \citep{perego14,fujibayashi17}.
Therefore, more realistic situation may be a combination of spatially
separated high, medium, and possibly low $\Ye$ components
as illustrated in Figure \ref{fig:schematic}.
In fact, the model with medium $\Ye$ does not perfectly reproduce
the flux at $< 5000$ \AA\ at $t = 2$ days
and the agreement can be improved with a presence of small
amount of high $\Ye$ ejecta probably near the pole.
Then, our line of sight may be somewhat off-axis
so that we can observe both high and medium $\Ye$ regions.
This may also explain the weakness of the gamma-ray emission
\citep{connaughton17,savchenko17,goldstein17,savchenko17_paper}.

Our interpretation implies that a large amount ejecta
with medium or high $\Ye$ is ejected during the post-merger phase.
The large ejecta mass suggests that the viscous mass ejection
is quite efficient in the NS merger event GW170817.
A required dimensionless viscous $\alpha$ parameter is $\alpha \gsim 0.03$
\citep{shibata17}.
In addition, we speculate that a relatively long-lived massive NS is
present after the merger \citep{metzger14,kasen15,lippuner17}
so that neutrino emission from the central NS can increase $\Ye$
of the surrounding disk as well as the ejecta from the disk.
It is, however, noted that self-consistent simulations from
merger to kilonova are necessary for drawing more quantitative conclusions.

\section{Conclusions}

We perform the radiative transfer simulations of kilonovae
for the optical and near-infrared counterpart of GW170817.
We show that the ejecta of 0.03 $\Msun$ containing lanthanide elements
explain the observed near-infrared emissions.
However the low-$\Ye$ component alone cannot explain
the optical emission at the initial phases.
As far as a single component model is considered,
the ejecta with medium $\Ye$ ($\sim 0.25$),
which contain a small fraction of lanthanide elements,
nicely explain both the optical and near-infrared emission simultaneously.
We suggest that the post-merger mass ejection plays
dominant contribution to the emission.

Our results have implications to the origin of $r$-process elements.
It is known that abundances of $r$-process elements in Galactic stars
show an universal pattern similar to those of the Sun \citep[\eg][]{sneden08},
which implies that a certain phenomenon produces
a wide range of $r$-process elements simultaneously.
The observed properties of SSS17a suggest that
the NS mergers eject not only dynamical ejecta
producing $r$-process elements beyond the second peak,
but also a substantial amount of post-merger ejecta with
medium or high $\Ye$ material producing elements
from the first to the second peaks.
Therefore, the NS mergers seem to synthesize
a full range of $r$-process elements \citep[\eg][]{just15,wu16}.
Mass ejection with an order of $0.01 \Msun$ is also enough
to explain the total amount of $r$-process elements in our Galaxy
\citep[\eg][]{piran14,hotokezaka15,rosswog17}
with the current estimates of the event rate.
These strengthen the hypothesis that the NS mergers are the origin of
$r$-process elements in the Universe.


\begin{ack}
We thank Kenta Hotokezaka and Shinya Wanajo for fruitful discussion.
Numerical simulations were carried out on the Cray XC30
at the Center for Computational Astrophysics,
National Astronomical Observatory of Japan.
This research was supported by the NINS program
for cross-disciplinary science study,
Inoue Science Research Award from Inoue Foundation for Science,
the research grant program of the Toyota Foundation (D11-R-0830),
the natural science grant of the Mitsubishi Foundation,
the research grant of the Yamada Science Foundation,
MEXT KAKENHI (JP17H06363, JP15H00788, JP24103003, JP10147207, JP10147214),
and JSPS KAKENHI (JP16H02183, JP15H02075, JP25800103, JP26800103).
The IRSF project is a collaboration between Nagoya University and the
South African Astronomical Observatory (SAAO) supported by
Optical \& Near-Infrared Astronomy
Inter-University Cooperation Program from MEXT of Japan
and the National Research Foundation (NRF) of South Africa.

\end{ack}

\end{document}